\documentclass[fleqn,usenatbib]{mnras}
\usepackage{newtxtext,newtxmath}
\usepackage[T1]{fontenc}
\usepackage{cancel}

\usepackage{graphicx}
\usepackage{amsmath}
\usepackage{bm}
\usepackage{xr-hyper}
\usepackage{xcolor}

\usepackage{bm,upgreek}
\usepackage{lineno}
\usepackage{soul}
\usepackage[normalem]{ulem}
\usepackage{caption}

\definecolor{mylinkcolor}{rgb}{0,0,0.8}
\definecolor{myurlcolor}{rgb}{0,0.5,0.5}
\definecolor{mycitecolor}{rgb}{0.5,0,0.5}

\allowdisplaybreaks

\title[density-split velocities]{Turning dispersion into signal: density-split analyses of pairwise velocities}
\author[Gon \& Cai]{\newauthor
Aritra Kumar Gon,\thanks{agon@roe.ac.uk}
Yan-Chuan Cai\thanks{cai@roe.ac.uk}
\\
Institute for Astronomy, University of Edinburgh, Royal Observatory, Blackford Hill, Edinburgh EH9 3HJ, United Kingdom\\
}

\date{Accepted 2026 April 27. Received 2026 April 6; in original form 2025 October 6}

\pubyear{\the\year}
\begin{document}
\label{firstpage}
% \pagerange{\pageref{firstpage}--\pageref{lastpage}}
\maketitle

\begin{abstract}
Pairwise velocities of the large-scale structure encode valuable information about the growth of structure. They can be observed indirectly through redshift-space distortions and the kinetic Sunyaev-Zeldovich effect. Whether it is Gaussian or non-Gaussian, the pairwise velocity has a broad distribution, but the cosmologically useful information lies primarily in the mean -the streaming velocities; the dispersion around the mean is often treated as a nuisance and marginalised over. This reduces the constraining power of our observations. We demonstrate that this is not necessarily the case, provided the underlying physics behind the dispersion is understood. By splitting the halo/galaxy samples according to their density environments and measuring the streaming velocities separately, the total signal-to-noise is several times greater than in conventional global measurements of the pairwise velocity distribution (PVD). This improvement arises because the global PVD is a composite of a series of near-Gaussian distributions with different means and dispersions, each determined by its local density environment. By splitting the data, we avoid cancellation between these opposing velocities, effectively turning the dispersion in the global PVD into a signal. Our findings indicate substantial potential for improving the analysis of PVD observations using the kinetic Sunyaev-Zeldovich effect and redshift-space distortions.
\end{abstract}


\section{Introduction}
The initial density perturbations of our Universe grow due to gravity, giving rise to peculiar velocities with respect to the Hubble flow. On large scales, peculiar velocities are connected to the model of the Universe and the law of gravity. In the linear regime and in Fourier space, $\bf v_k$ is mapped to the matter density contrast $\delta_{\bf k}$ through the continuity equation, ${\bf v_k}= -iafH \delta_{\bf k} {\bf k}/k^2$,  where $a$ is the scale factor, $f=d\ln{D}/d\ln{a}\approx \Omega_m^{\gamma}$ is the linear growth rate, with $D$ being the linear growth parameter and $\Omega_m$ the matter density parameter; $H$ is the Hubble constant at $a$. In general relativity (GR), $ \gamma \approx 0.55$ \citep{Peebles1980, Lahav1991} but $\gamma$ can have different values if the law of gravity is non-GR \citep[e.g.][]{Linder2007,Guzzo2008}. Therefore, peculiar velocities encode information about the growth of structure; they can be used to test the law of gravity. 

In observations, peculiar velocities are often measured in pairs, called the pairwise velocities $v^p$, defined as the difference of velocity vectors for a pair of galaxies (or dark matter halos) along their separation vector $\mathbf{r}$ in 3D. 

When observed with a sample of galaxies (or halos), the mean of the pairwise velocity, called the mean streaming velocity, is the ensemble average of the number-weighted $v^p$ \citep{Juszkiewicz_Springel_Durrer1999,Bernardeau2002,Gon2024}, 
\begin{align}
\label{expect_val_density_weighted}
\bar{v}^{p}(r)=\frac{\Big\langle \left(\mathbf{v_1}-\mathbf{v_2}\right)\cdot \mathbf{\hat{r}}\,(1+\delta^h_1)(1+\delta^h_2)\Big\rangle}{\Big\langle (1+\delta^h_1)(1+\delta^h_2)\Big\rangle},
\end{align}
where $\delta^h_1$ and $\delta^h_2$ are the galaxy (or halo) density contrast at $\bf r_1$ and $\bf r_2$ respectively, and $\mathbf{\hat{r}} = ({\mathbf{r_1}-\mathbf{r_2}})/{|\mathbf{r_1}-\mathbf{r_2}|}$. In the linear regime and for tracers with the linear bias b,  Eq.~\ref{expect_val_density_weighted} is reduced to
\begin{align}
\label{expect_fin1} 
\bar{v}^{p}(r)=-\frac{2}{3}\frac{[afH](z)\,r\,	b \bar{\xi}(r)}{1+\,b^2\xi(r)},
\end{align}
where
$\bar{\xi}(r)=\frac{3}{r^3}\int_{0}^{r} r'^2\xi(r')dr'$ \citep[e.g.][]{Juszkiewicz_Springel_Durrer1999, Ferreira1999, Sheth2001} (for a derivation, see the Appendix-A of \citep{Gon2024}). 
where
$\bar{\xi}(r)=\frac{3}{r^3}\int_{0}^{r} r'^2\xi(r')dr'$ \citep[e.g.][]{Juszkiewicz_Springel_Durrer1999, Ferreira1999, Sheth2001} (for a derivation, see the Appendix-A of \citep{Gon2024}). The mean streaming velocity is uniquely coupled to the two-point correlation function (2PCF) of matter, $\xi(r)$. It is related to cosmology through those coupling parameters. It is clear that in the linear regime, $v^p$ has a dispersion determined by the dispersion of $\xi(r)$. However, non-linear growth of peculiar velocities, particularly shell-crossing, breaks the above mapping and increases the velocity dispersion. In principle, the full pairwise velocity distribution (PVD), $P(v^p)$, is also related to the underlying cosmological model, but the connection is less explicit. In addition, the PVD is typically non-Gaussian \citep[e.g.][]{Efstathiou1988, Zurek1994,Juszkiewicz1998,Scoccimarro2004}, which is challenging to model. Although there has been significant effort in modelling the velocity dispersion \citep[e.g.][]{Sheth1996, Sheth_Hui2001, Kuwabara2002, Seto1998, Scoccimarro2004, Slosar2006, Valogiannis2020}, in observational analyses, it is often treated as a nuisance -- a component to be marginalised over during the fit in order to extract cosmological information from the streaming velocity. In observations, this means that one has to average over a large sample of galaxy pairs to beat the dispersion. Such is the case for the pairwise kinetic Sunyaev-Zeldovich (kSZ) estimator \citep[e.g.][]{Ferreira1999} and the empirical non-linear models of redshift-space distortions (RSD) \citep[e.g.][]{Peacock2001, Guzzo2008}. However, by marginalising over the nuisance parameters related to velocity dispersions, we weaken the constraining power for cosmology from our data. We will show that this does not have to be the case with a different way of analysing the data. 
\begin{figure*}
\hspace{-0.5cm}\includegraphics[width=0.5\linewidth]
{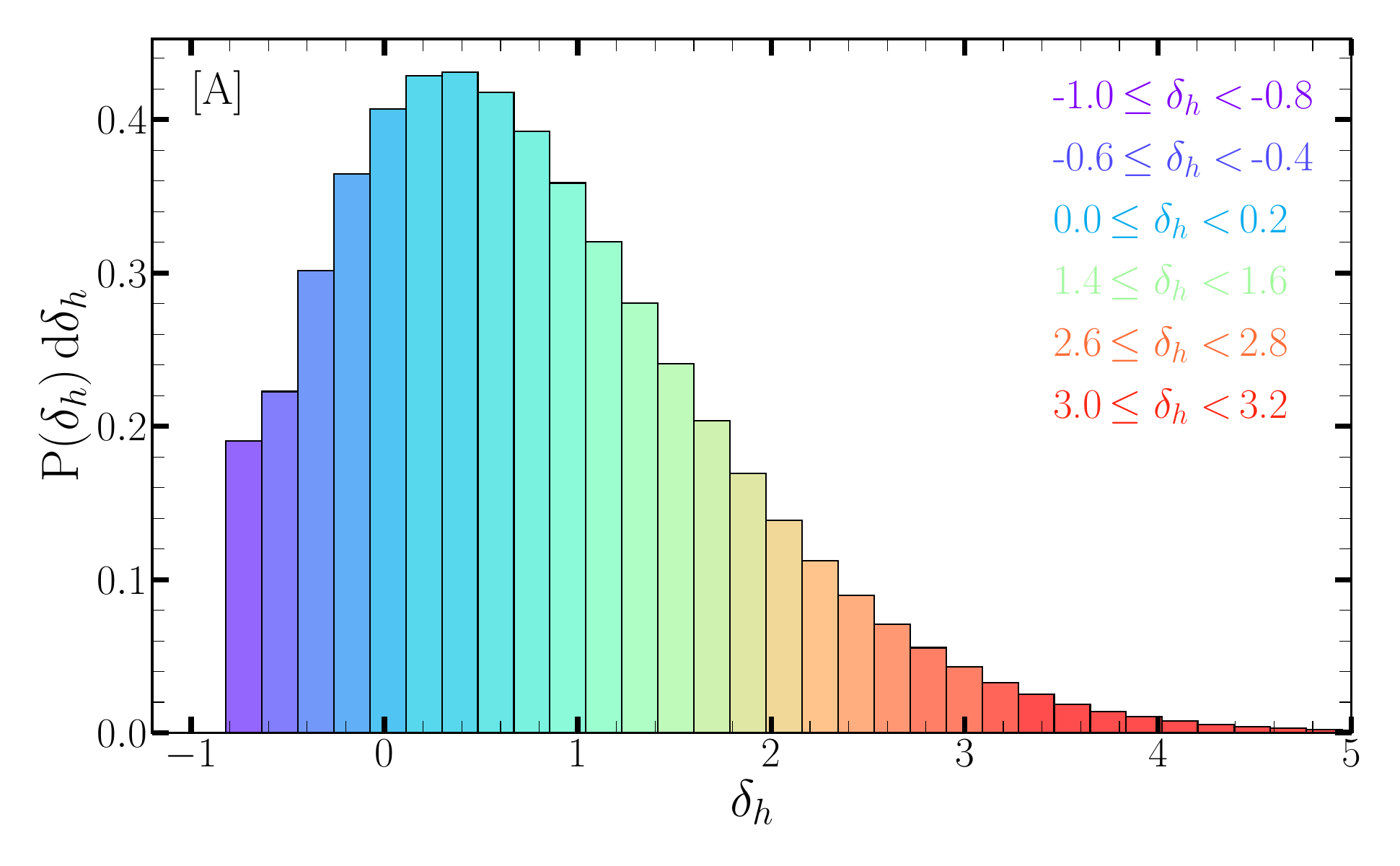}
\hspace{-0.3cm}\includegraphics[width=0.5\linewidth]
{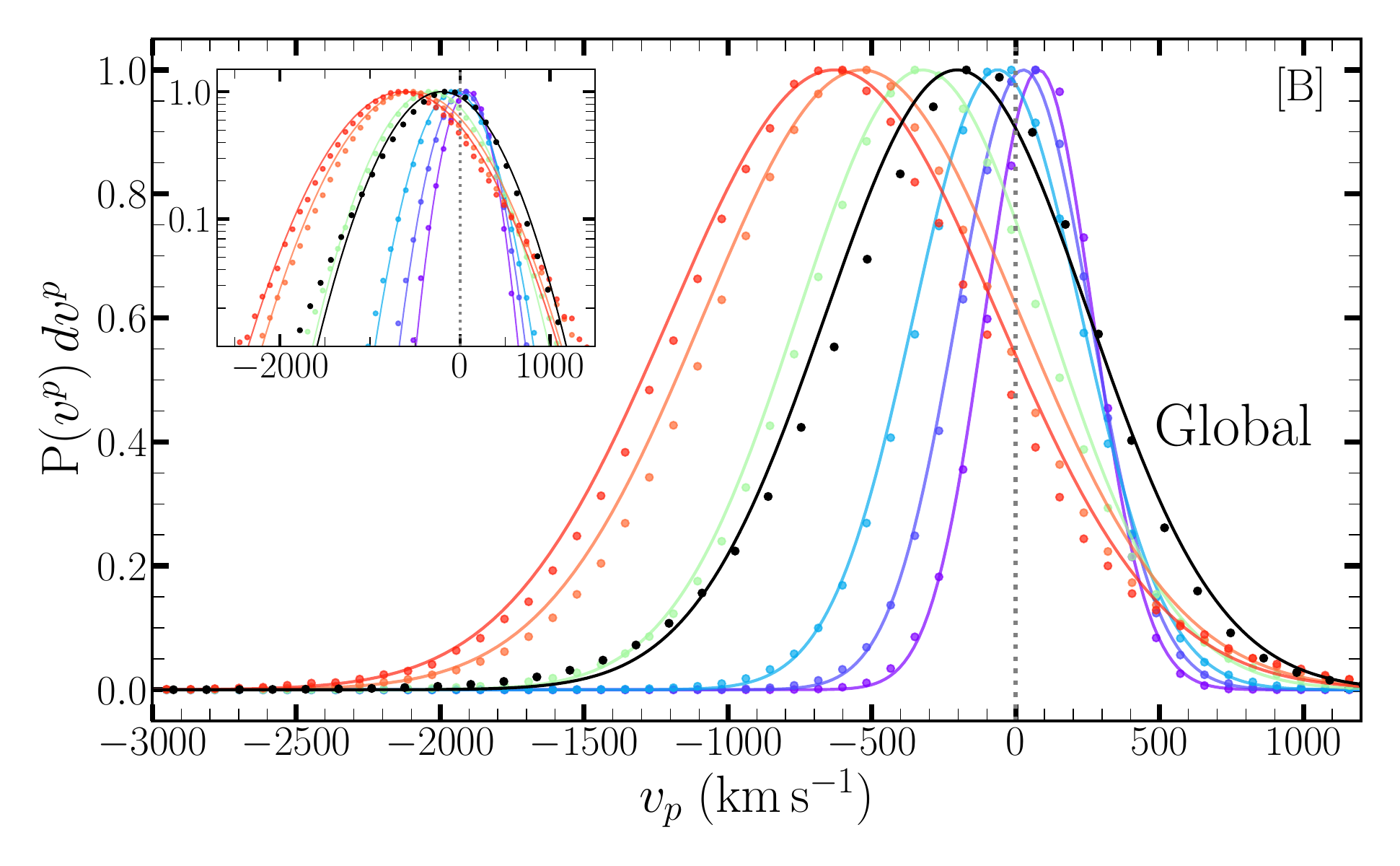}
\caption{Panel-A: the probability distribution of the overdensity $\delta_h$ at the locations of halos from the Quijote simulation at $z=0$. $\delta_h$ is smoothed by a top-hat window function with the radius $R_{\rm sm}=25\,\mathrm{Mpc\,h^{-1}}$. The density bins are shown in different colours, with some of them indicated by the legend with matched colours. These are the environmental densities used for Panel-B. Panel-B: The pairwise velocity distributions in different density environments, shown in matched colours to the legend of Panel-A. Dots are measurements from the simulation, lines are the best-fit Gaussian functions, with the best-fit parameters shown in Table~\ref{tab:tab2}. The global PVD for halos in the entire simulation box, labelled as `global', is shown in black for comparison. The inset shows the version with a logarithmic scale for the $y$-axis. }
\label{fig:pairwise_density_and_vel_distribution}
\end{figure*}

In observations, peculiar velocities can be measured indirectly via the following two ways:\\
(1) redshift-space distortions --- The observed line-of-sight distance of a galaxy is perturbed by its peculiar velocity. This induces anisotropies to the observed clustering of galaxies, known as redshift-space distortions (RSD) \citep{Kaiser1987}. In Fourier space and in the linear regime, the redshift-space power spectrum is mapped to the real-space version via $P^s({\bf k})=(1+\mu^2 f)^2 P^r({k})$, where $\mu=\cos\theta$ and $\theta$ is the angle subtended from the line of sight \citep{Kaiser1987}. From Eq.~\ref{expect_fin1}, $f$ is linearly mapped to $\bar{v}^p$ and so only the streaming velocity enters the modelling, i.e., any dispersions of the peculiar velocity are essentially noise to be averaged out (see the empirical model in \citep[e.g.][]{Peacock2001} where the dispersions are included and marginalized over).
The problem of dealing with velocity dispersions is also seen in configuration space. The redshift-space correlation function $\xi(s, \mu)$ is mapped to the real-space version $\xi(r)$ via $1+\xi(s, \mu)=\int dv_{\parallel}[1+\xi(r)]p(v_{\parallel},r)$ \citep{Peebles1980}, where $p(v_{\parallel},r)$ is the line-of-sight pairwise velocity distribution function for galaxies at the 3D distance separation of $r$. Predictions from perturbation theories do well in capturing the shape of $p(v_{\parallel},r)$, but its absolute amplitude is usually accounted for by a free parameter \citep[e.g.][]{Reid2011,Wang2014, Vlah2016}, and so marginalisation over it is needed in observational analyses. \\

(2) The kinetic Sunyaev-Zeldovich effect --- The coherent line-of-sight peculiar motions of ionised gas associated with galaxies, groups, and clusters alter the frequency of CMB photons scattered by them through the Doppler effect. This increases/decreases the line-of-sight CMB temperature if the ionised gas is moving towards/away from the observer, known as the kinetic Sunyaev-Zeldovich (kSZ) effect \citep{Sunyaev_Zeldovich1972, Sunyaev1980}. 
The pairwise kSZ estimator is commonly used for measuring the large-scale velocity correlations, which is proportional to the streaming velocity and therefore the global 2PCF \citep[e.g.][]{Ferreira1999, Gon2024}. Here, like RSD, the dispersion of the PVD is also treated as noise \citep[e.g.][]{Hand2012}.   

In summary, the PVD plays a central role in RSD and kSZ. In both cases, often the streaming velocity of the PVD contributes to the model prediction, and the dispersions are treated as a nuisance. One way to exploit the PVD for more cosmological information is to go beyond the conventional two-point statistics, particularly to decompose the full PVD according to local properties of the large-scale structure. There have been indications that analysing the observed density field by splitting it into different environments warrants more cosmological information than the conventional two-point statistics \citep{Gruen2018,Enrique2021, Repp2021}. One possible reason for the improvement may come from sampling the non-Gaussian probability density function (PDF) of the density \citep{Enrique2021, Paillas2023, Giblin2023}. Following the same spirit, in this letter, we will perform density-split analysis for the velocity field. 

\section{Splitting the pairwise velocity distribution} 

Previous studies have shown that PVD depends on the large-scale local density \citep[e.g.][]{Tinker2007, Hellwing2016}. 
\citep{Hellwing2016} showed that the PVD varies significantly in different cosmic web environments. \citep{Tinker2007} developed a model based on the density-dependence of PVD for modelling RSD in terms of the two-point correlation function, but the regime of $\delta<0$ was not explored; \citep{Enrique2021} applied the density-split (DS) technique to expand it to $\delta<0$ for the velocity PDF, but has not looked at the PVD. We will generalise the DS technique to analyse the PVD across all density environments. We use the {\sc rockstar} halo catalog from the Quijote low-resolution simulations \citep{Quijote_sims} for the analysis. Here are the major steps we take to perform density-split analyses of the PVD:\\
\\
(1) We smooth the halo field with a top-hat window of the radius $R_{\rm sm}=25\,\mathrm{Mpc\,h^{-1}}$, and measure the PDF of the density field {\it at the locations of halos}, shown in the left-hand panel of Fig.~\ref{fig:pairwise_density_and_vel_distribution}. In this way, we associate each halo with a value of $\delta_h$, representing the density environment in which they are situated. \\
\\
(2) We then split $\delta_h$ into density bins as indicated by different colors in Fig.~\ref{fig:pairwise_density_and_vel_distribution}. It is clear that the PDF of $\delta_h$ is non-Gaussian at the smoothing scale of $25\,\mathrm{Mpc\,h^{-1}}$, with its peak skewed towards positive values of $\delta_h$ and the tail of the distribution extended to large values. Note that the PDF of the density field sampled at the locations of halos is not the same as the density PDFs sampled at random (or equally spaced) locations, usually referred to as counts in cells, see \citep[e.g.][]{Peebles1980, Uhlemann2016, Repp2021, Enrique2021}.\\
\\
(3) We measure the PVD for pairs of halos in different density bins and at different separation scales $r$, noted as $P(v^p, r|\delta_h)$. We can see $P(v^p, r|\delta_h)$ as the conditional PVD, conditioned at $\delta_h$. An example for $25\,\mathrm{Mpc\,h^{-1}} < r <30\,\mathrm{Mpc\,h^{-1}}$ is shown in the right-hand panel of Fig.~ \ref{fig:pairwise_density_and_vel_distribution}. The global pairwise velocity distribution measured with all halos in all environments, noted as $P(v^p,r)$, is shown for comparison\footnote{Note that the sum of $P(v^p, r|\delta_h)$ over all bins of $\delta_h$ does not equal to the global PVD i.e., $P(v^p,r) \neq \int P(v^p, r|\delta_h) d \delta_h$, because $P(v^p,r)$ contain pairs crossing different $\delta_h's$. We will get back to this point in the latter part of this letter. }. We find that even though the global PVD is non-Gaussian, $P(v^p, r|\delta_h)$ in different $\delta_h$ environments is well-fit by a Gaussian function of different means and dispersions. Both the mean $\bar v^p(r | \delta_h)$ and the dispersion $\sigma^p_v(r | \delta_h)$ are strongly correlated with the local density $\delta_h$ (see Fig.~\ref{fig:sigma_skewness} and Table~\ref{tab:tab2}). 

\begin{figure}
\hspace{-.5cm}
\includegraphics[width=0.50\textwidth]
{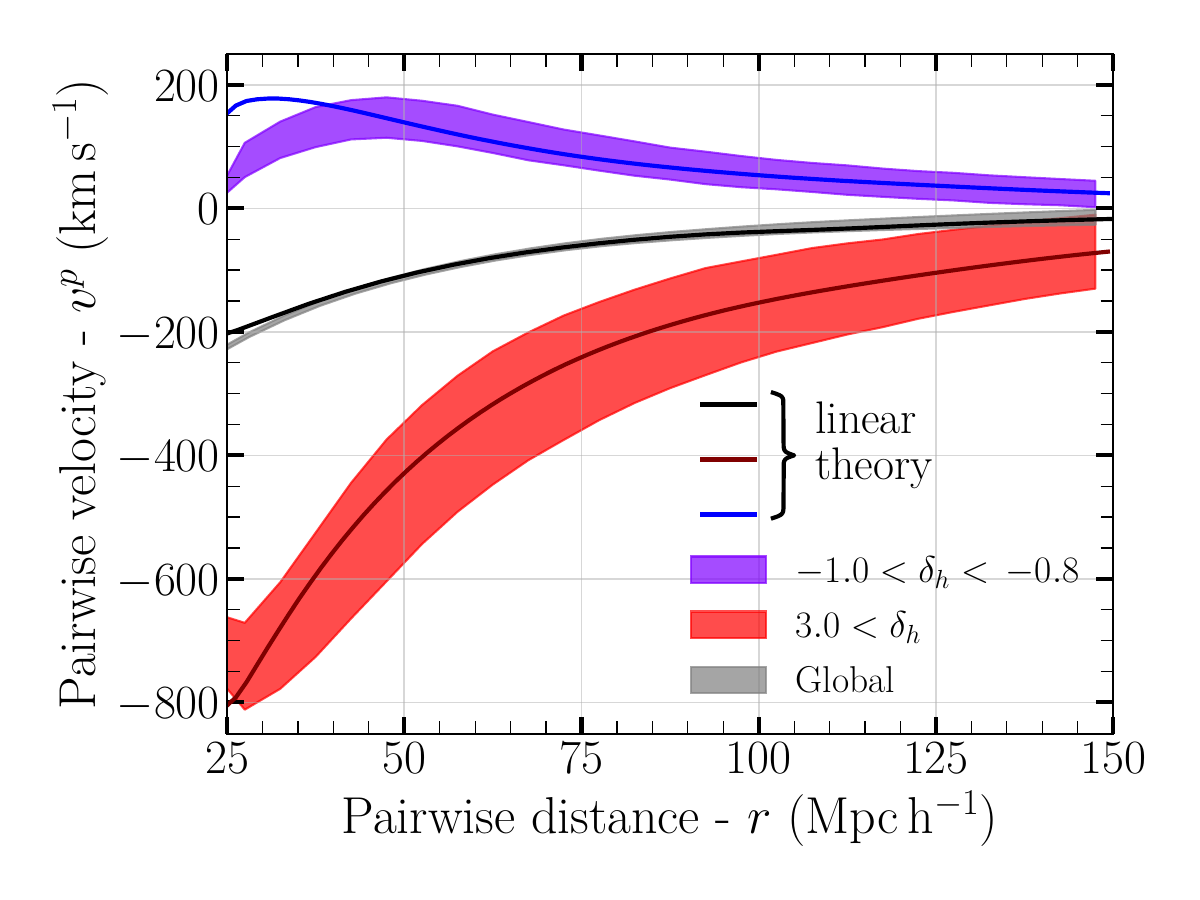}
\caption{ 
The mean pairwise velocity around different density environments (shown in different colours) versus the pair-distance separation $r$. The shaded colour regions are 1-$\sigma$ errors corresponding to the sample of halos in the 1-Gpc$~h^{-1}$-aside Quijote simulation. The values of the environmental density are indicated by the legend. For comparison, the global mean pairwise velocity is shown in grey. Lines are linear theory predictions using the linear matter correlation function, a linear halo bias $b$ and linear density-split bias $b_{\rm DS}$ (in the DS cases, Eq.~\ref{expect_fin2}). The bias parameters are measured directly from the Quijote simulations. The agreement between the lines and the shaded regions is within 20-30\% for most scales shown.}
\label{fig:pairwise_vel_vs_dist_high_low_box}
\end{figure}

For the mean, $\bar v^p(r | \delta_h)$, it is negative (or positive) when $\delta_h <0$ (or $\delta_h >0$), implying that, on average, halos are moving apart from (or towards) each other when they are in an underdense (or overdense) region.  Fig.~\ref{fig:pairwise_vel_vs_dist_high_low_box} gives examples of the $\bar v^p(r | \delta_h)$ for an underdense $(\delta_h<-0.2)$ and an overdense $(\delta_h>3.0)$ region respectively, in comparison with the global one $\bar v^p(r)$. We can see that $\bar v^p(r | \delta_h)$ remains positive (or negative) out to large scales, indicating that the underdensities and overdensities preserve their signs from $25\,\mathrm{Mpc\,h^{-1}}$ to large scales. The average pairwise outflow velocity around underdense regions peaks at around 200~km/s, and the infall velocity peaks at approximately 800~km/s, which is four times the amplitude of the global mean pairwise velocity. In addition, $\bar v^p(r | \delta_h)$ and $\bar{\xi}(r | \delta_h)$ are linearly coupled over a wide range of scales as shown in Fig.~\ref{fig:pairwise_vel_vs_dist_high_low_box}. 
Note that in the DS case, the velocity fields are now conditioned on our selected low or high-density regions. Thus, the linear theory prediction for the mean pairwise velocity is the correlation between the conditional velocity field and the entire halo field. So, another linear bias factor $b_{\rm DS}$ related to the density splitting should enter \citep{Xu2024}, and Eq.~\ref{expect_fin1} becomes
\begin{align}
\label{expect_fin2} 
\bar{v}^{p}(r)=-\frac{2}{3}\frac{[afH](z)\,r\, b_{\rm DS}\,b \,\bar{\xi}(r)}{1+\,b^2\xi(r)}.
\end{align}
With the bias parameters fixed from the Quijote simulations, the above linear theory prediction agrees with direction within 20-30\% for most scales shown.
For the dispersion, $\sigma^p_v(r | \delta_h)$, it increases with $\delta_h$, as seen in the right-hand panel of Fig.~\ref{fig:pairwise_density_and_vel_distribution}. Their relationship is well described by the empirical formula $\sigma^p_v(r| \delta_h)=\sigma^p_{vA}(r)(1+\delta_h)^\alpha + \sigma^p_{vB}$, shown in Fig.~\ref{fig:sigma_skewness} and Table~\ref{tab:tab2}. We find $\alpha=1$ at $r=27.5$~$\mathrm{Mpc\,h^{-1}}$ (see also \citep{Sheth2001,Hamana2003}.) We also find the skewness $\mu_3(r|\delta_h)$ of the PVD is well fit by a fourth-order polynomial function (Table~\ref{tab:tab2}). Its absolute value is minimised at approximately $\delta_h\sim 1.5$. It becomes negative with decreasing $\delta_h$. 

\subsection{Splitting to avoid cancellation}
The above analyses suggest that the global non-Gaussian pairwise velocities distribution can be decomposed into a series of narrow conditional PVDs, each having its own local mean. Another way to see this is that the broadness of the global PVD is partly due to the sum of all halo pairs in a mixture of density environments. Once the pairs of halos are split according to their environments, we are able to reveal a spectrum of coherent outflow and infall velocities, each of them having a relatively small dispersion. Averaging the global distribution yields a single coherent streaming velocity corresponding to the global clustering, or 2PCF -- at the cost of cancelling the positive and negative streaming velocities between the low and high dense regions, yielding a large dispersion. By splitting, however, we recover the multiple components of outflows and infalls. We are effectively turning the noise -- the {\it global} velocity dispersion, into signal -- the {\it local} means of positive or negative velocities. This has the following implications:\\
\\
(1) Models of observables of the PVD, such as kSZ and RSD, usually have an explicit dependence on the mean PVD, but not on the dispersion, and so the DS-technique allows us to make full use of the coherent velocities hidden in the environmental-dependence of the PVD. Crucially, each bin of the local streaming velocity is approximately linearly coupled to the conditional correlation function (Fig.~\ref{fig:pairwise_vel_vs_dist_high_low_box}), and so the cosmological information contained in them is explicit and straightforward to track. \\
\\
(2) We see that the amplitudes of the local mean streaming can be significantly greater than the global mean around high-density regions (the red curve in Fig.~\ref{fig:pairwise_vel_vs_dist_high_low_box}), and that the dispersions are significantly smaller in low-density regions (blue and purple curves in the right-hand panel of Fig.~\ref{fig:pairwise_density_and_vel_distribution}). We can therefore anticipate a higher signal-to-noise for pairwise-velocity analyses using the DS technique than the conventional one. This is what we show next.

\begin{figure}
\includegraphics[width=0.5\textwidth]{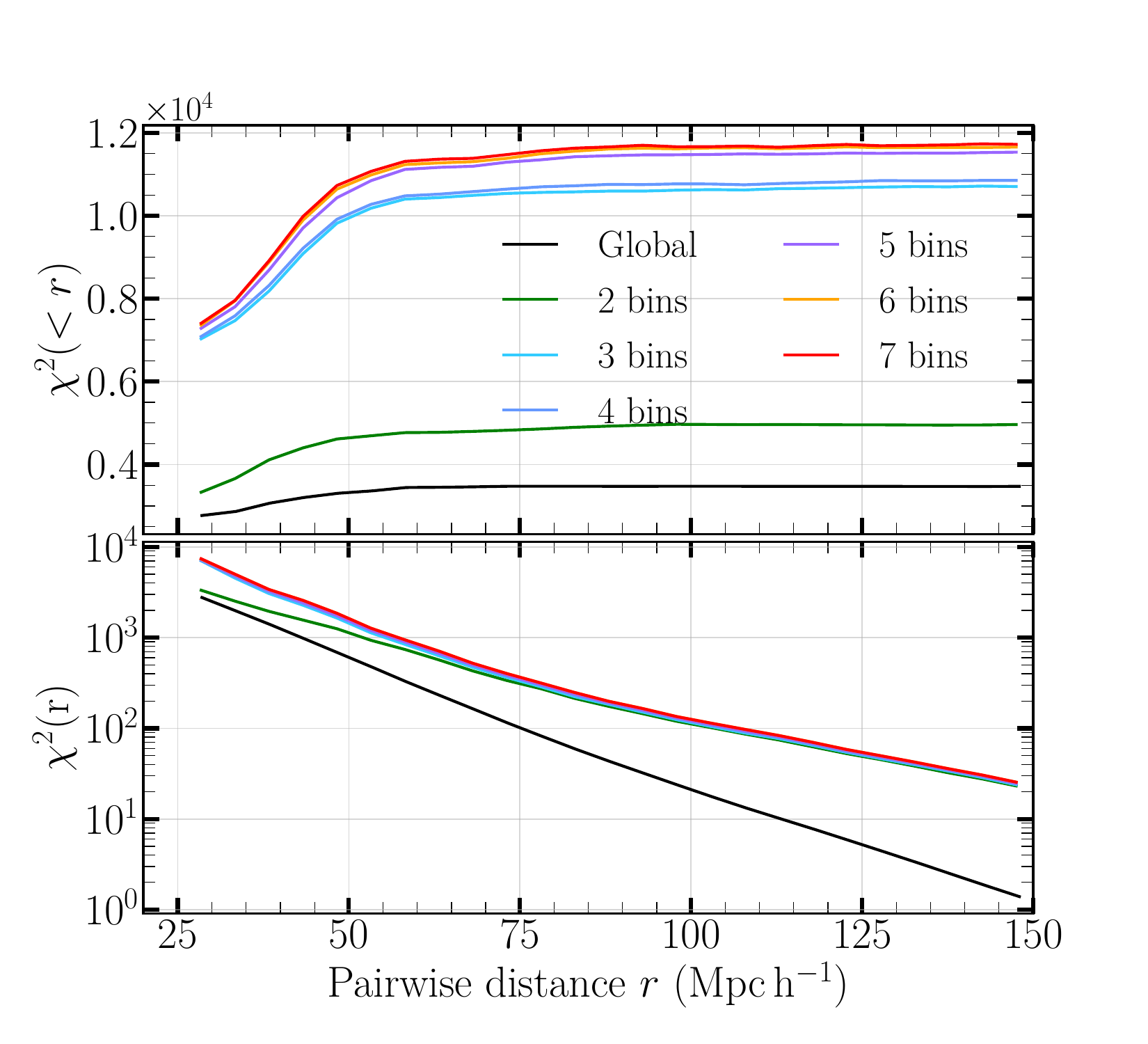}
\caption{\label{fig:chi_sq} Top-panel: The cumulative $\chi^2$, characterising the deviation of the mean pairwise velocity from null, measured with the density-split technique and the conventional pair-counting (Global). The $\chi^2$ values are cumulative with respect to the pair separation $r$, starting from 25~$\mathrm{Mpc\,h^{-1}}$ with the bin-width of 5~$\mathrm{Mpc\,h^{-1}}$. It is clear that the smallest scale radial bin contributes more than half of the signal, as expected for two-point functions.
Measurements with the DS method with different numbers of density bins are shown in different colours. Lower-panel: 
Similar to the top panel but showing the differential $\chi^2$ i.e., the $\chi^2$ per radial bin of 5-$\mathrm{Mpc\,h^{-1}}$ width. }
\end{figure}
$\,$

\subsection{Increased signal-to-noise} 

To compare the information contained in the PVD using the DS technique versus the global pair-counting of the 2PCF method, we use $\chi^2$ to quantify the significance of the deviation of a measurement from null,
%\begin{equation}
$\chi^2=\bar v^p \,C^{-1}\, (\bar v^p)^{T}.$
%\end{equation}
All variables on the right-hand side of the equation are functions of $r$, with the matrix $C$ accounting for the covariance among different $r$-bins. We are interested in the differential and cumulative $\chi^2$ up to a maximum $r$ for the separation distances between $25\,\mathrm{Mpc\,h^{-1}}$ to $150\,\mathrm{Mpc\,h^{-1}}$, with the bin-width of $5\,\mathrm{Mpc\,h^{-1}}$. The lower bound is set to be the smoothing radius, and the choice of the upper bound has little effect on the total signal-to-noise. 

We measure the mean pairwise velocities with 4000 realisations of the simulated halos using the DS and 2PCF methods, respectively, and treat them as data vectors. The covariance matrix {\bf $C$} is constructed using the same 4000 simulations. We have checked for the convergence of our results by increasing the number of simulations gradually from 1000 to 4000 and confirm that the $\chi^2$ has converged with this number of simulations.

To make sure that we maximise the information with the DS technique, there are subtleties about the binning related to the DS technique. First, how many density bins should we split? For this, we explore splitting the halo density field progressively from 2 bins to 7 bins, and find that the total $\chi^2$ has reached a good level of convergence by 6 bins (Fig.~\ref{fig:chi_sq}). Note that the exact number of bins by which the result converges may also depend on the splitting thresholds, which we do not explore in this paper.
Second, for DS, counting all halo pairs within each density bin is incomplete because there are pairs whose two halos belong to different density bins. Missing these cross-bin pairs in the pair counting will lead to a loss of information. For a split with $N$ bins, there are $\,^NC_2$ cross bins. We therefore include all the cross-bin pairs in our data vector to ensure that the total number of pairs in DS adds up exactly the same as in the 2PCF method. In this way, the length of the data vector is $N(N+1)/2$ for each $r$. 

We see from the top panel of Fig.~\ref{fig:chi_sq} that for all cases, there is an increase in $\chi^2$ from $25\,\mathrm{Mpc\,h^{-1}}$ to $\sim 50\,\mathrm{Mpc\,h^{-1}}$, beyond which the $\chi^2$ plateaus. This implies that most of the information comes from relatively small scales, as also shown in the lower panel. Comparing $\chi^2$ for the global PVD of the conventional 2PCF, there is an increase of a factor of 1.4, 3.1, and 3.4 with the DS-technique when we have 2, 3, and 6 density bins, respectively. From the $\chi^2$ for each individual $r$-bin in the lower panel of Fig.~\ref{fig:chi_sq}, we find that the $\chi^2$ with the DS-technique is at least 2$\times$ greater than the global case on small scales. The factor of gain increases with scale, reaching 10$\times$ at 150~$\mathrm{Mpc\,h^{-1}}$. 

The gain of information with the DS-technique on large scales suggests that non-Gaussianity is not a necessary condition for the DS-method to be effective, as the PVD on such large scales is much closer to Gaussian (see Fig.~\ref{fig:PDF_LargeScale} in supplementary material~\ref{Apendix}). In fact, for a near-Gaussian distribution, it is typically more symmetric than a highly non-Gaussian one, and thus, the cancellation between positive and negative pairwise velocities will be more severe in the global case. Therefore, we anticipate a more significant improvement with density-split towards the linear Gaussian regime.

The global PVD there can also be further decomposed into local Gaussian PDFs with narrower widths (Fig.~\ref{fig:PDF_LargeScale}). Perhaps more importantly, on those large scales (r $\gtrsim$ 80~$\mathrm{Mpc\,h^{-1}}$), DS with two bins is able to maximise the gain of information --  no need for splitting into any more bins. This supports our reasoning earlier that the global streaming velocity suffers severe cancellations when averaging over the global PVD, whereas with the DS method, splitting into two bins of positive and negative local streaming velocities is sufficient to avoid cancellations, thus turning the global noise into signal.

For the density field, it is well recognised that two-point statistics are sufficient to extract all the information when the field is Gaussian. Our results suggest that this may depend on how the analysis is conducted. The evidence of the extra information from the DS method is obvious in the bottom panel of Fig.~\ref{fig:chi_sq}. We suspect that the extra information may arise partly from the non-perfect Gaussianity of the field even on large scales, due to discrete sampling. Perhaps more relevant, the velocities on split fields are conditional quantities. There is an implicit correlation between the split density field and the velocity field. This could bring information from different scales, and is different from the global two-point method. In addition, most observables in cosmology are in redshift space, which means that the influence of the velocity field and therefore scale/mode mixing is unavoidable. This suggests potential for gaining more information with the DS method on large scales.

\section{Discussions and Conclusions}
\subsection{Observational implications} 
{The significant increase in the signal-to-noise for the pairwise velocities measured using the density split technique over the conventional 2PCF suggests great potential for improving cosmological constraints using observations from galaxy redshift surveys and CMB experiments. Analyses of pairwise kSZ and RSD will benefit from the DS method. 

The mean streaming velocity is proportional to the pairwise kSZ signal, and we have shown that it is positive and negative for underdense and overdense density environments, respectively. This implies that we can observe pairwise kSZ signals of the opposite sign using the density split technique. This information is not present in the global pair counting, i.e., when averaging all pairs of galaxies over the entire survey volume. With the DS method, we can exploit the cosmological information from the pairwise kSZ signal in different density regions separately, which, as we have shown with simulations, has the potential to lead to a higher signal-to-noise for kSZ measurements using the same data. For real kSZ measurements, one must also include other sources of noise, such as the instrument noise or noise coming from the primary CMB fluctuations, apart from the statistical noise, which we have included in this analysis. An application to observational data will be presented in a separate paper.} 

For redshift-space distortions, the significant benefit of DS analysis against 2PCF has been demonstrated in \citep{Enrique2021, Paillas2023}, but the reason for the improved constraints on cosmology has not been entirely clear. It was speculated to be mainly due to the sampling of the non-Gaussian PDF of the density field, which is likely to be partly true. Now we can see that another reason must be the better sampling of the mean streaming velocities in different environments using the DS technique. 

Perhaps more importantly, the way we conduct density split in this letter is subtly different from what was applied in \citep{Gruen2018, Friedrich2018, Enrique2021,Paillas2023}, and more similar to the case presented in \citep{Abbas2005, Tinker2007}, or \citep{Chiang2014, Repp2022} in Fourier space. The density PDF in this work was evaluated at the locations of tracers (halos), rather than at random locations, or uniform sampling, as in \citep{Uhlemann2016, Gruen2018, Friedrich2018, Enrique2021,Repp2021,Paillas2023}. This subtlety has significant implications for the modelling of RSD. As seen in DS-RSD (or void-RSD), the random sampling of DS (or void) centres in redshift space becomes non-random in real space, and vice versa. This poses challenges for modelling DS-RSD (or void-RSD), because of the non-trivial mapping for the DS (or void) centres between real and redshift space
\citep{Nadathur2019a}. An iterative reconstruction was developed to bring the tracer field back to the real space to split the density \citep[e.g.][]{Nadathur2019b}; forward modelling with simulations \citep{Cuesta-Lazaro2024, Paillas2024} was shown to be successful to some extent. With this version of density-split, we always have halos at the centres where the density PDF was evaluated. Our pair-counting method is therefore the same as the 2PCF except that halo pairs are re-grouped according to their local densities. This makes the modelling of our DS-RSD no fundamentally different from 2PCF. The extra factor we need to take care of is the bias associated with environmental dependence, which on scales of tens of $\mathrm{Mpc\,h^{-1}}$, is approximately linear \citep{Repp2021, Xu2024}. Still, this is slightly entangled with the fact that the densities are evaluated in redshift-space for the split, which we will show is not significant in a separate paper. With the new method, perturbation theories that work for 2PCF should work (even better) for DS-clustering. With this, both the modelling and the actual data analysis can be aligned.

\section{data availability}
The Quijote simulations \citep{Quijote_sims} used in this paper are public. Our codes and results are available upon reasonable request to the authors.

\section{Acknowledgments}
We thank John Peacock, Andrew Repp, Finn Roper, Ravi Sheth, Istvan Szapudi, and Cora Uhlemann for useful discussions. We thank the anonymous referee for providing very useful comments. YC acknowledges the support of the UK Royal Society through a University Research Fellowship. This research was partly supported by the Munich Institute for Astro-, Particle and BioPhysics (MIAPbP), which is funded by the Deutsche Forschungsgemeinschaft (DFG, German Research Foundation) under Germany´s Excellence Strategy – EXC-2094 – 390783311. For the purpose of open access, the author has applied a Creative Commons Attribution
(CC BY) license to any Author Accepted Manuscript version arising from this submission.

\bibliographystyle{mnras}
\bibliography{references}% Produces the bibliography via BibTeX.
%%%%%%%%%%%%%%%%%%%%%%%%%%%%%%%%%%%%%%%%%%%

%%%%%%%%%%%%%%%%% APPENDICES %%%%%%%%%%%%%%%%%%%%%
% \newpage
\appendix
% \clearpage

\section{Mean, dispersion, and skewness of velocity pdfs at different smoothing scales}
\label{Apendix}

\begin{figure}
\includegraphics[width=1.\linewidth]{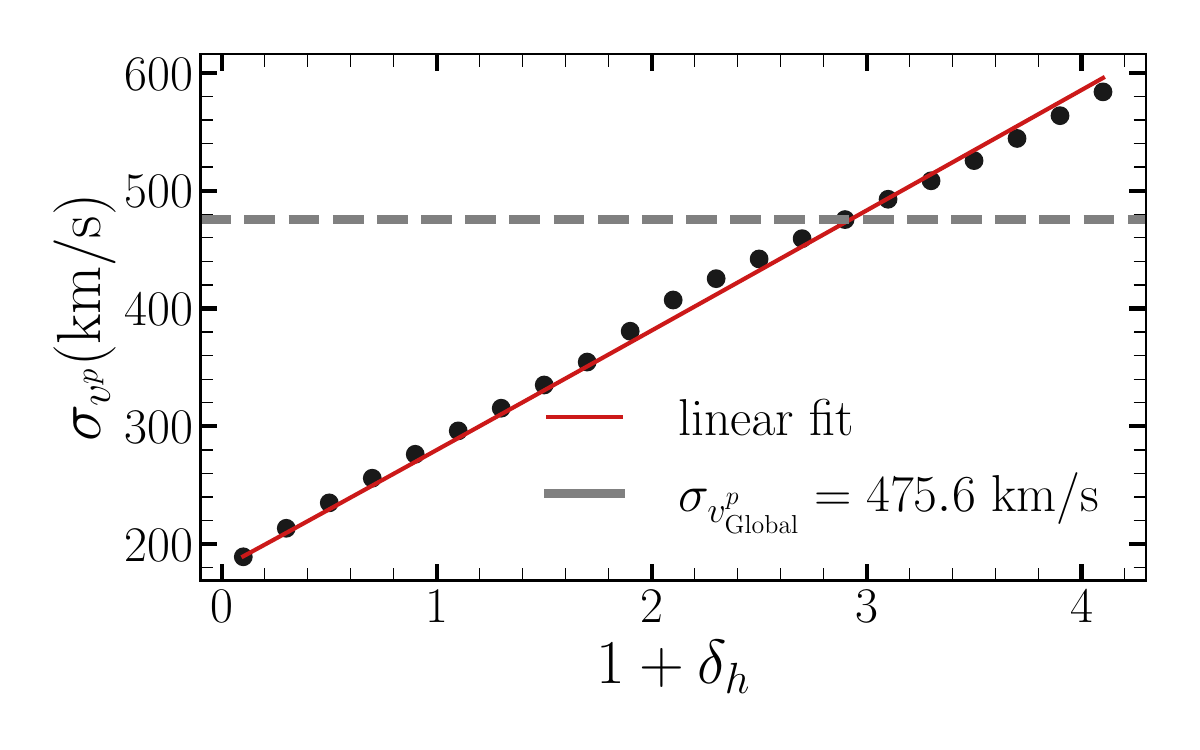}

\includegraphics[width=1.\linewidth]{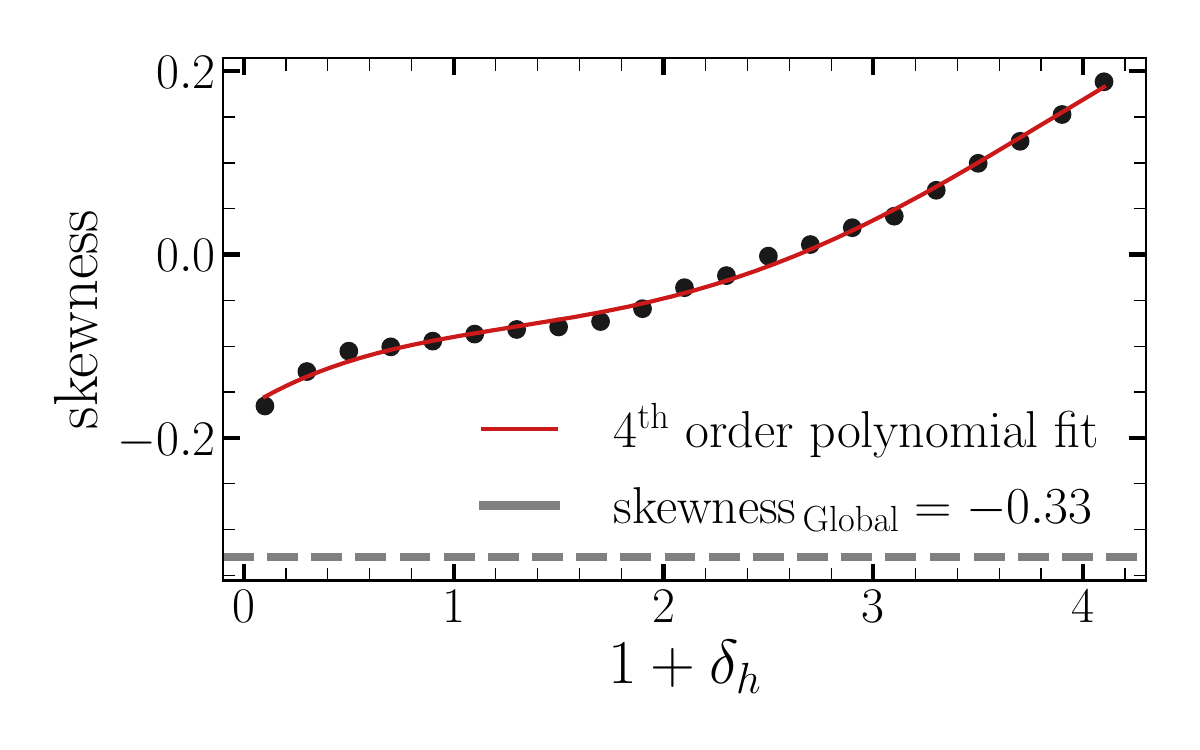}
\vspace{-0.5 cm}
\caption{Left: the best-fit function, $\sigma_v^p/\mathrm{(km/s)}=100(1+\delta_h)+180$, for the pairwise velocity dispersions versus environmental density (1+$\delta_h$). The data points from the simulation are measured for the environmental densities matched to those shown in Panel-A of Fig.~\ref{fig:pairwise_density_and_vel_distribution}. The dispersion of the global pairwise velocity distribution of the entire halo population is shown in the grey-dashed line, labelled as $\sigma_{v^p_{\rm Global}}$.  Right: similar to the left, but showing measurements of the skewness as well as the best-fit results with a fourth-order polynomial function (skewness = $-0.0032\,\delta^4_h +0.0198\,\delta^3_h -0.0152 \,\delta^2_h + 0.0404 \,\delta_h -0.0898$). The global skewness of the PVD of the entire halo population, skewness$\,_{\rm Global}$, is shown in a grey-dashed line. The absolute value of skewness$\,_{\rm Global}$ is significantly larger than those in specific density environments. This suggests that the PVDs in specific density environments are more Gaussian than the global case.}
\label{fig:sigma_skewness}
\end{figure}

\begin{table}
\centering
% \captionsetup{width=\linewidth} 
\begin{tabular}{p{2.5cm} p{0.9cm} p{0.9cm} p{0.9cm} p{0.9cm}}%{lcccc}
\hline
Sample                 & mean (km\,s$^{-1}$) & dispersion (km\,s$^{-1}$) & skewness & kurtosis \\
\hline
Global                 & -209.6 & 475.6 & -0.33  & 0.763 \\
$-1.0 < \delta_h < -0.8$ & 79.0   & 189.3 & -0.16  & 0.192 \\
$-0.6 < \delta_h < -0.4$ & 25.9   & 235.1 & -0.10  & 0.505 \\
$0.0 < \delta_h < 0.2$   & -63.8  & 296.1 & -0.08  & 0.615 \\
$1.4 < \delta_h < 1.6$   & -323.3 & 442.0 & -0.001 & 0.617 \\
$2.6 < \delta_h < 2.8$   & -537.8 & 544.2 & 0.12   & 0.698 \\
$3.0 < \delta_h < 3.2$   & -623.7 & 583.7 & 0.18   & 0.811 \\
\hline
\end{tabular}
\caption{\label{tab:tab2}The mean, dispersion, skewness, and kurtosis measured from the pairwise velocity distributions in different density environments, $\delta_h$, indicated in the first column. These environment densities are matched to those shown in the legend of Panel-A of Fig.~\ref{fig:pairwise_density_and_vel_distribution}, and so the pairwise velocity distributions are matched to those in Panel-B of Fig.~\ref{fig:pairwise_density_and_vel_distribution}. The row `Global' represents the results measured from the pairwise velocity distribution of the entire halo population. }
% \label{tab:tab2}
\end{table}

\begin{figure*}
\includegraphics[width=0.47\textwidth]{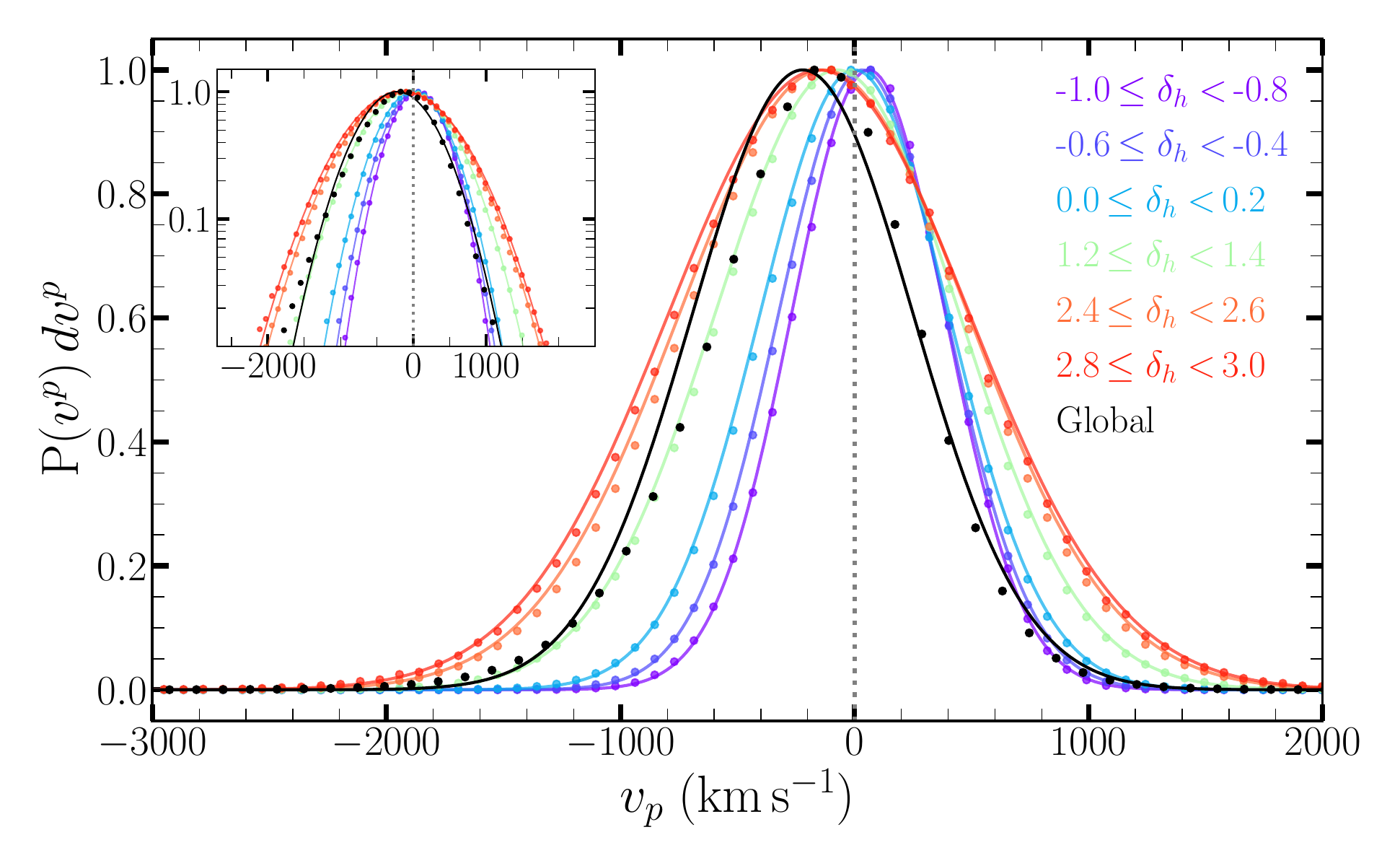}%
\includegraphics[width=0.47\textwidth]{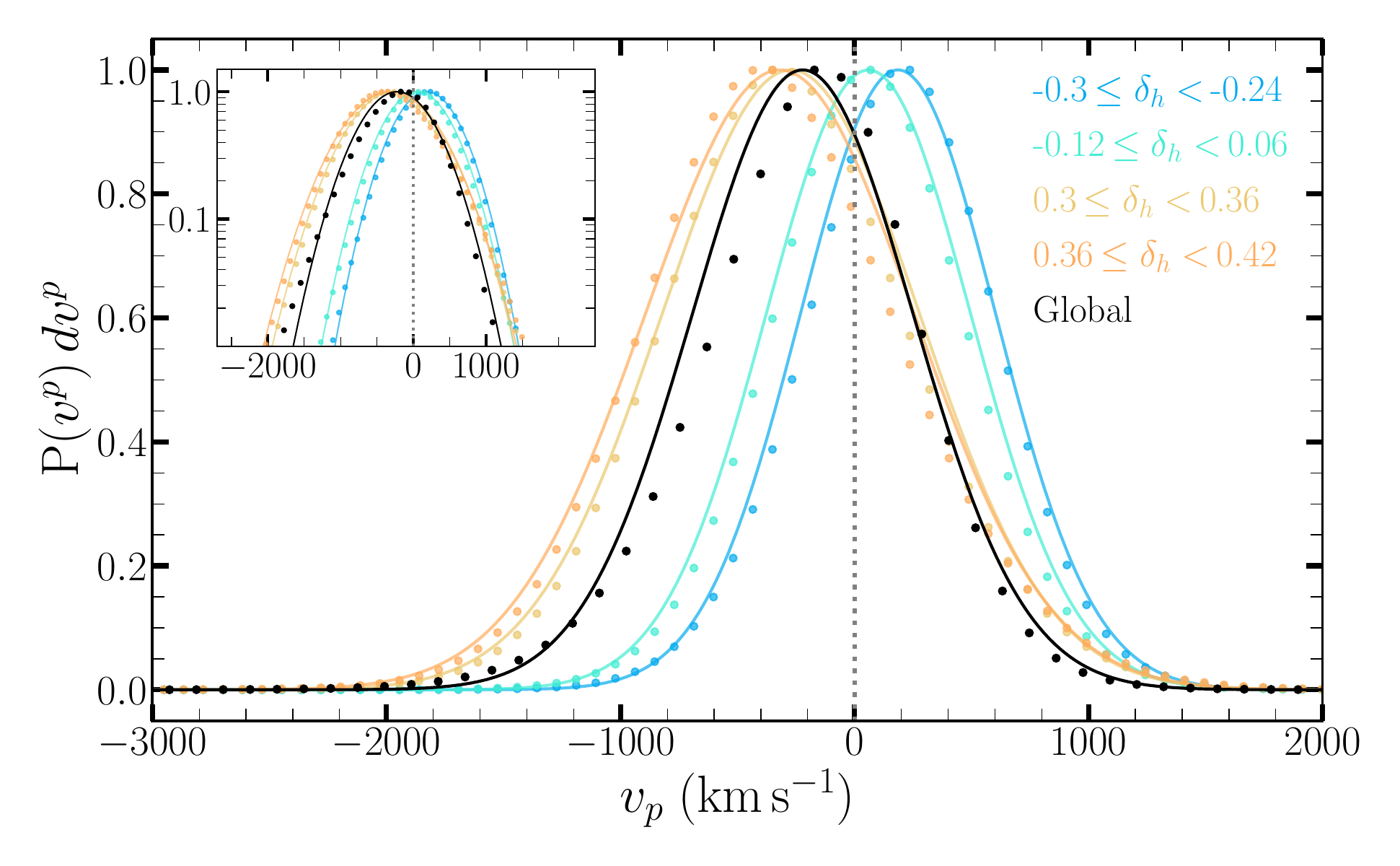}
\vspace{-0.5 cm}
\caption{The pairwise velocity distribution of simulated halos measured at the scale of $\sim$ 95~$\mathrm{Mpc\,h^{-1}}$. Different colours indicate different environmental densities. Upper Panel: the environmental densities, labelled as $\delta_h$, are measured and split at 25~$\mathrm{Mpc\,h^{-1}}$, but the pairwise velocity distribution are measured at 95~$\mathrm{Mpc\,h^{-1}}$. Lower Panel: both the environmental densities and the PVD are measured at 95~$\mathrm{Mpc\,h^{-1}}$. `Global' represents the pairwise velocity distribution of the entire halo population. Compared to Panel-B of Fig.~\ref{fig:pairwise_density_and_vel_distribution}, which represents PVDs at the scale of 25~$\mathrm{Mpc\,h^{-1}}$, these large-scale PVDs are relatively more Gaussian and with their peaks being less offset from zero.}
\label{fig:PDF_LargeScale}
\end{figure*}

\end{document}